\documentclass[letter,twocolumn]{jpsj2}
\usepackage{graphicx}

\def\runtitle{
Theoretical Study on Resonant Inelastic X-Ray Scattering 
in Quasi-One-Dimensional Cuprates}
\def\runauthor{Takuji \textsc{Nomura} and Jun-ichi \textsc{Igarashi}}

\title{
Theoretical Study on Resonant Inelastic X-Ray Scattering 
in Quasi-One-Dimensional Cuprates}
\author{Takuji \textsc{Nomura}\thanks{E-mail: nomurat@spring8.or.jp} 
and Jun-ichi \textsc{Igarashi}}
\inst{Synchrotron Radiation Research Center,
Japan Atomic Energy Research Institute,
Mikazuki, Sayo, Hyogo  679-5148}
\recdate
{
}
\abst{
We study theoretically the resonant inelastic x-ray scattering 
in quasi-one-dimensional insulating 
copper oxides, where the incident photon energy 
is tuned to the Cu1$s$-4$p$ absorption energy. 
Our attention is focused particularly 
on the strong momentum-transfer dependence 
of the spectral shape observed in recent experiments. 
We describe the antiferromagnetic ground state 
within the Hartree-Fock theory, and consider charge excitations 
from the ground state within the random phase approximation. 
By taking account of the electron correlation effects perturbatively, 
we obtain detailed momentum-transfer dependence of the spectra 
in a semiquantitative agreement with the experiments.}

\kword{Resonant inelastic x-ray scattering, One-dimensional copper oxide, 
Sr$_2$CuO$_3$, SrCuO$_2$}
\begin{document}
\maketitle

Resonant inelastic x-ray scattering (RIXS) measurements
are now becoming a promising and unique experimental technique 
to clarify the detailed charge excitation spectra of solids 
in a relatively high-energy range~\cite{Rf:Kotani2001},
owing to high brilliance of synchrotron radiation.
In particular, the RIXS measurements in the hard x-ray regime could be
a new powerful tool to detect momentum dependent charge excitations in solids.
A number of compounds have been investigated by this technique
~\cite{Rf:Kao1996,Rf:Hill1998,Rf:Abbamonte1999,Rf:Hasan2000,
Rf:Hasan2002,Rf:Kim2002,Rf:Inami2003,Rf:Kim2003c}. 
Among them, the RIXS in cuprates utilizing the Cu1$s$-4$p$ 
absorption edge attracts much interest recently, 
since relatively large momentum dependence was indeed observed in the charge excitation spectra
~\cite{Rf:Hill1998,Rf:Abbamonte1999,Rf:Hasan2000,Rf:Hasan2002,Rf:Kim2002,Rf:Kim2003c}. 

\begin{figure}
\begin{center}
\includegraphics[width=60mm]{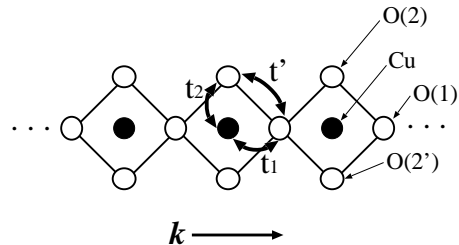}
\end{center}
\caption{Schematic figure of the Cu-O chain 
presenting the definition of the transfer integrals.}
\label{Fig:CuOchain}
\end{figure}

The electronic properties of the quasi-one-dimensional (Q1D) 
insulating cuprates Sr$_2$CuO$_3$ (SCO213) and SrCuO$_2$ (SCO112)
are well characterized by the one-dimensional chain consisting of the Cu-O plaquettes. 
The Cu-O chains are expected only weakly coupled with each other~\cite{Rf:Motoyama1996}. 
In SCO213 the Cu-O plaquettes are aligned 
with sharing the corner oxygen atoms with the nearest plaquettes, 
as shown in Fig.~\ref{Fig:CuOchain}. 
It was reported that SCO213 shows antiferromagnetic order 
below about 5 K with reduced staggered magnetization along the chain. 
The strong fluctuations are responsible 
for the extremely reduced N\'eel temperature 
and magnetization~\cite{Rf:Kojima1997}, 
but the ground state of SCO213 is an antiferromagnetic (AF) insulator. 
Concerning SCO112, the Cu-O plaquettes form a zigzag chain 
by sharing the edges with the nearest plaquettes. 
The zigzag chain can be regarded approximately as being constructed 
by combining two independent Cu-O chains sharing 
the edges with each other~\cite{Rf:Motoyama1996}. 
Therefore we study the RIXS properties of these two cuprates 
by considering the corner-sharing Cu-O chain as in Fig.~\ref{Fig:CuOchain} 
in the present article. 

Recently, RIXS measurements were performed 
for the Q1D cuprates SCO213~\cite{Rf:Hasan2002} 
and SCO112~\cite{Rf:Hasan2002,Rf:Kim2003c}. 
In the both Q1D cuprates, characteristic spectral weight 
was obtained for energy transfer $\omega \approx 2, 5.6$ (eV), 
but the origin of the spectral peaks are still unclear. 
The aim of the present article is to provide a quantitative 
understanding on the RIXS, particularly on the position of the peaks 
and the highly momentum-dependent behaviors of the spectra 
in these one-dimensional cuprates. 
We calculate the intensity in the RIXS spectra 
as a function of energy-momentum transfer, 
using the perturbation theory developed 
by Platzman and Isaacs~\cite{Rf:Platzman1998}. 
We describe the antiferromagnetic (AF) ground state 
by the Hartree-Fock (HF) theory, 
and take account of the excitations from the ground state 
within the random phase approximation (RPA). 
As a result, we obtain a semiquantitative agreement 
with experiments, regarding the peak positions and 
the momentum dependent behaviors. 

The essential microscopic process of the RIXS in the cuprates 
is considered to be as follows~\cite{Rf:Tsutsui2000}. 
The incident photon excites resonantly the Cu1$s$ core electrons 
to the upper empty Cu4$p$ bands. 
Since the core Cu1$s$ orbitals are well localized 
to be coupled strongly to the Cu3$d$ electrons 
by the interorbital Coulomb interaction, 
the created Cu1$s$ core hole scatters the Cu3$d$ electrons, 
which occupy the bands relatively close to the chemical potential
(As a result of the scattering, 
the most probably only one electron-hole pair is created on the Cu3$d$ band). 
In order to describe this process specifically, 
we consider the total Hamiltonian of the form, 
$H=H_{dp}+H_{1s-3d}+H_{1s}+H_{4p}+H_x$. 
The $dp$-Hamiltonian $H_{dp}$ describing the electronic properties 
of the Cu3$d$-O2$p$ bands is given in the form, $H_{dp}=H_0+H'$. 
$H_0$ and $H'$ are the noninteracting and the interaction parts, respectively. 
\begin{equation}
H_0 = \sum_{k\sigma} 
\mib{d}_{k\sigma}^{\dag} \left(
\begin{array}{@{\,}cccc@{\,}}
\varepsilon_d & \zeta_1(k) & t_2 & - t_2 \\
- \zeta_1(k) & \varepsilon_{p_1} 
& \zeta'(k) & - \zeta'(k) \\
t_2 & - \zeta'(k) & \varepsilon_{p_2} & 0 \\
- t_2 & \zeta'(k) & 0 & \varepsilon_{p_2} 
\end{array}
\right)\mib{d}_{k\sigma}, 
\end{equation}
where $\mib{d}_{k\sigma} = (d_{k\sigma},p_{1k\sigma},p_{2k\sigma},p_{2'k\sigma})$, 
and $d_{k\sigma}$, $p_{1k\sigma}$, $p_{2k\sigma}$ and $p_{2'k\sigma}$ 
are the electron annihilation operators 
in the atomic $d_{x^2-y^2}$ or $p_{x,y}$ orbits 
at the Cu, O(1), O(2) and O(2') atoms, respectively. 
The dispersions are $\zeta_1(k) = 2{\rm i}t_1\sin \frac{k}{2}$ 
and $\zeta'(k) = 2{\rm i}t'\sin \frac{k}{2}$. 
The hopping parameters are 
$t_1=-1.45$ eV, $t_2=-1.8$ eV and $t'=-0.7$ eV (See Fig.~\ref{Fig:CuOchain}), 
as determined for SCO213 
by the LDA first principle calculation~\cite{Rf:Neudert2000}. 
The interacting part is 
\begin{equation}
H'= \frac{U_d}{N} \sum_{kk'q} d_{k-q \uparrow}^{\dag} 
d_{k'+q \downarrow}^{\dag}  d_{k' \downarrow} d_{k \uparrow}. 
\end{equation}
The Coulomb energy is taken to be $U_d=11$ eV 
in the present study. Regarding the one-particle level, 
we take $\varepsilon_{p_2} - \varepsilon_{p_1} = 0.5$ eV, 
$\varepsilon_d^{\rm HF} - \varepsilon_{p_1} = -0.5$ eV, 
where $\varepsilon_d^{\rm HF}$ is the screened 
Cu3$d$ one-electron energy within the HF theory. 
For this parameters, we obtain the AF ground state 
with the staggered magnetization $m_{\rm stagg.} = 0.43 \mu_{\rm B}$. 
The scattering of the Cu3$d$ electrons 
by the Cu1$s$ core hole is described by 
\begin{equation}
H_{1s-3d} = \frac{V}{N} \sum_{kk'q\sigma\sigma'}
d^{\dag}_{k'+q\sigma} s^{\dag}_{k-q\sigma'}
s_{k\sigma'} d_{{k'}\sigma},
\end{equation}
where $s_{k\sigma}$ ($s^{\dag}_{k\sigma}$) 
is the annihilation (creation) operator for the Cu1$s$ electrons 
with momentum $k$ and spin $\sigma$, and 
$V$ is the so-called core-hole potential. 
$H_{4p}$ and $H_{1s}$ describe the kinetics 
of the electrons on the Cu4$p$ and Cu1$s$ bands.
Since the Cu1$s$ electrons are well localized, we take 
completely flat dispersion for them. 
For the Cu4$p$ electrons, we use simple two-dimensional cosine-shaped 
band for simplicity. This simplification does not affect the spectral 
shape so drastically, because the factor containing the Cu4$p$ dispersion function
is integrated up in momentum, as seen later in eq.~(\ref{Eq:intensity}). 
$H_x$ describes the transitions between the Cu1$s$ and Cu4$p$ states, 
involving the photon absorption and inverse emission processes.
$H_x$ is of the form
\begin{equation}
H_x =\sum_{kq\sigma} (w(q; k)
p'^{\dag}_{k+q\,\sigma} s_{k\,\sigma} + {\rm h.c.}),
\end{equation}
where $p'^{\dag}$ is the creation operator of the Cu4$p$ electron.
In the present study, we ignore the momentum dependence
of the matrix elements  $w(q; k)$, i.e., $w( q; k)=w$.

\begin{figure}
\begin{center}
\includegraphics[width=55mm]{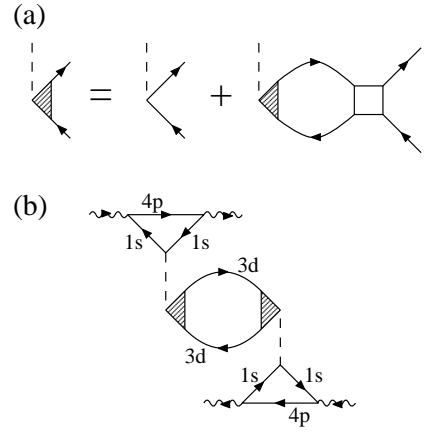}
\end{center}
\caption{
(a) Vertex renormalization in the RPA. 
The solid lines, the shaded triangle and the empty square 
are the Green's function for Cu3$d$ electrons, the renormalized vertex 
and the antisymmetrized bare Coulomb interaction, respectively.
(b) The diagrammatic representation of the RIXS intensity.
The wavy and the dashed lines denote photons and the 1$s$-core-hole 
potential, respectively.
The Green's functions for Cu3$d$, Cu1$s$ and Cu4$p$ electrons 
are assigned to the solid lines marked with '3d', '1s' and '4p', 
respectively.}
\label{Fig:diagrams}
\end{figure}
The main spectral weight of the RIXS is evaluated 
by the diagram shown in Fig.~\ref{Fig:diagrams}. 
In Fig.~\ref{Fig:diagrams}(a), the effective three point vertex 
function $\Lambda(\omega,q)$ represented by the shaded triangle 
is calculated within the RPA, 
and this vertex part is inserted 
to the diagram Fig.~\ref{Fig:diagrams}(b) 
representing the total scattering probability.
In Fig.~\ref{Fig:diagrams}(b), note that the off-diagonal components of 
the Keldysh Green's functions are assigned to the solid lines which connect 
the upper normally-time-ordered branch and the lower reversely-time-ordered branch, 
while the usual causal (normally-time-ordered) Green's functions are assigned to the solid lines 
in the upper branch and the reversely-time-ordered Green's functions are assigned 
to the solid lines in the lower branch~\cite{Rf:Nozieres1974}. 
Regarding the core hole potential $V$, 
we take only the first order term in $V$ (i.e., Born scattering). 
The effect of higher orders in $V$ is briefly mentioned later. 
The analytic expression for the diagram Fig.~\ref{Fig:diagrams}(b) is obtained as 
{\scriptsize
\begin{eqnarray}
W(\omega_i q_i; \omega_f q_f) = (2\pi)^3 N |w|^4 \sum_{k jj'}
\delta (E_j(k)+\omega-E_{j'}(k+q)) \nonumber\\
\times n_j(k)(1-n_{j'}(k+q)) \biggl| \sum_{\sigma\sigma'}\
U_{j, d\sigma}(k) \Lambda_{\sigma\sigma'}(\omega,q) 
U_{d \sigma', j'}^{\dag}(k+q) \nonumber\\
\times \sum_{k_1}
\frac{V}{(\omega_i+\varepsilon_{1s}+i\Gamma_{1s}-\varepsilon_{4p}(k_1))
(\omega_f+\varepsilon_{1s}+i\Gamma_{1s}-\varepsilon_{4p}(k_1))}
\biggr|^2,
\label{Eq:intensity}
\end{eqnarray}}
where $q_i$ and $q_f$ ($\omega_i$ and $\omega_f$) are the momenta (energies) 
of the initially absorbed and the finally emitted photons, respectively. 
$q = q_i - q_f $ and $\omega = \omega_i - \omega_f $ are the momentum transfer 
and the energy loss, respectively. 
$E_j(k)$ and $n_{j}(k)$ are the energy dispersion and the electron occupation number 
of the band $j$, respectively, obtained by diagonalizing the HF Hamiltonian for $H_{dp}$. 
$U_{j, d \sigma}(k)$ is the $(j, d \sigma)$-element of the unitary matrix 
diagonalizing the HF Hamiltonian for $H_{dp}$.
$\varepsilon_{1s}$ and $\varepsilon_{4p}(k)$ are the kinetic energies 
of the Cu1$s$ and Cu4$p$ electrons, respectively. 
In the present study the incident photon energy $\omega_i$ is tuned 
to the Cu1$s$-4$p$ absorption energy 
$\omega_i \approx \varepsilon_{4p}(0) - \varepsilon_{1s}$, 
for which the intensity of the spectrum is enhanced by the resonance. 
The decay rate $\Gamma_{1s}$ of the core hole is 0.8 eV in the present study. 

\begin{figure}
\begin{center}
\includegraphics[width=75mm]{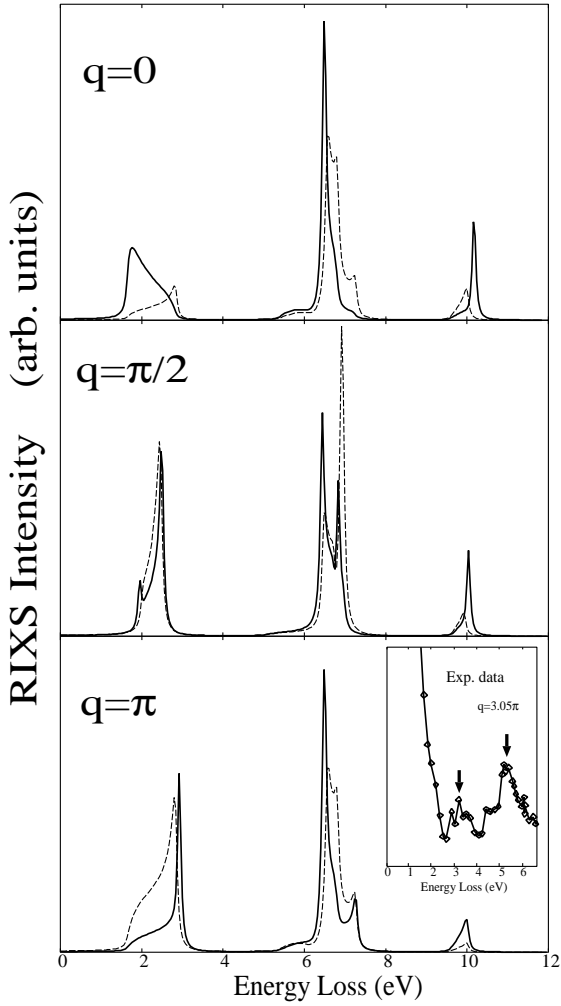}
\end{center}
\caption{The calculated spectra as a function of energy loss 
for three momentum transfers $q=0, 0.5 \pi$ and $\pi$.
The thick solid and the thin broken lines are the results 
with and without the RPA corrections, respectively.
The inset shows a typical experimental spectrum read 
from ref.~\ref{Rf:Hasan2002}, 
and the arrows point the two peaks.}
\label{Fig:1ddpsp}
\end{figure}
\begin{figure}
\begin{center}
\includegraphics[width=75mm]{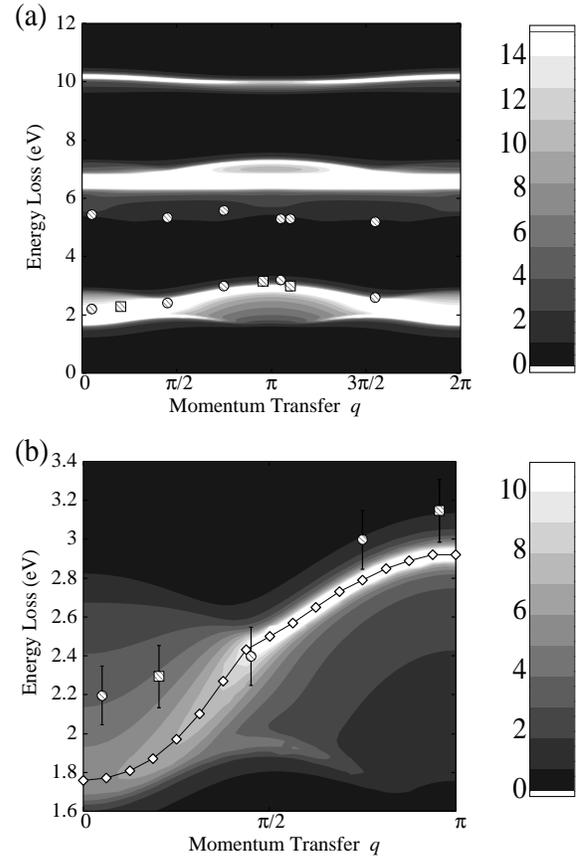}
\end{center}
\caption{
Graylevel plots for the spectra as a function 
of momentum transfer and energy loss. 
The light and dark regions correspond 
to high and low RIXS intensity, respectively. 
The hatched squares and circles are experimental peak positions 
in SCO213 and SCO112, respectively, from ref.~\ref{Rf:Hasan2002}. 
In (b), the diamonds connected by a thin line 
denote the theoretical peak positions.}
\label{Fig:1ddpspgl}
\end{figure}

The numerical results of the RIXS intensity by the formula eq.~(\ref{Eq:intensity})
are shown as a function of energy loss $\omega$ for momentum transfers 
$q=0, \frac{\pi}{2}$ and $\pi$ in Fig.~\ref{Fig:1ddpsp}. 
We see that there are three characteristic peaks in the calculated RIXS 
spectra around $\omega \approx 2, 6, 10$ eV. 
In the experiments, two peaks are observed around 
the momentum transfer $\omega \approx 2$ and 5.6 eV, 
which are in agreement with the present numerical results. 
The high-energy 10eV peak gives only small 
contribution to the total spectral weight, 
and the intensity around $\omega \approx 2$ and 6 eV 
covers the main part of the spectral weight. 
In Fig.~\ref{Fig:1ddpspgl}, the contourplots of the calculated RIXS spectra 
are shown as a function of the momentum transfer $q$ and the energy loss $\omega$, 
and the experimental data read from ref.~\ref{Rf:Hasan2002} 
are also displayed for comparison. 
We can see that the 2eV peak shows a relatively large sinusoidal dispersion, 
while the other two do not. 
This sinusoidal dispersion of the 2eV peak quantitatively 
agrees with the experiments~\cite{Rf:Hasan2002,Rf:Kim2003c}. 
Concerning the 6eV peak, the calculated peak positions deviate 
somewhat from the experimental ones, maybe due to the crude 
tight-binding fitting. The weak momentum dependence 
of the 6eV peak position is well reproduced by the calculation. 

\begin{figure}
\begin{center}
\includegraphics[width=80mm]{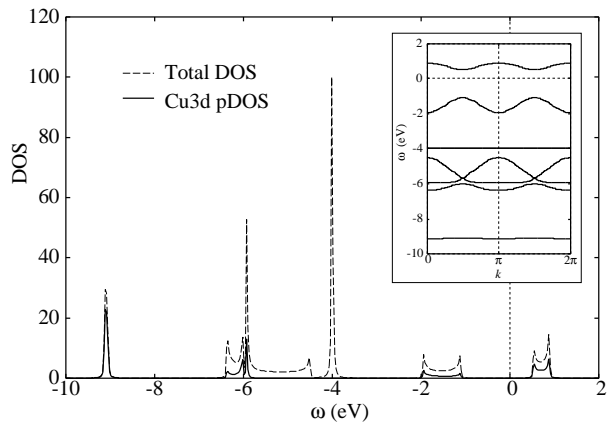}
\end{center}
\caption{The density of states for the antiferromagnetic ground state. 
The thin dotted and thick solid lines represent the total 
and the Cu3$d$ partial density of states, respectively. 
The chemical potential is set to $\omega=0$.
The inset shows the band dispersions.}
\label{Fig:dos}
\end{figure}

In order to inspect the origin of the peaks in the RIXS spectra, 
we should turn back to the formula (\ref{Eq:intensity}). 
Comparing the results modified by the correlations 
(the solid lines in Fig.~\ref{Fig:1ddpsp})
with the uncorrected results
(the dashed lines in Fig.~\ref{Fig:1ddpsp}), 
we can see that the overall spectral shape is 
roughly captured even without any correlation effects, 
although the detailed spectral structure are of course 
affected by the correlation effects. 
Therefore we might expect that the overall spectral shape is determined 
by the other factors rather than the correlation effects on the scattering process,
and thus expect naturally that the factor $n_j(k)(1-n_{j'}(k+q))
|U_{j, d\sigma}(k) U_{d\sigma', j'}^{\dag}(k+q)|^2$ is essential  
for determining the overall spectral shape. 
It is very interesting to note that this factor is the product 
of the partial electron occupation number of the band $j$ at the momentum $k$
[ $n_j(k)|U_{j, d\sigma}(k)|^2$ ]
and the partial hole occupation number of the band $j'$ with the momentum $k+q$
[ $(1-n_{j'}(k+q))|U_{d\sigma', j'}^{\dag}(k+q)|^2$ ].
Thus the Cu1$s$-4$p$ RIXS spectra in cuprates are closely related 
to the partial occupation number of the Cu3$d$ electrons in each band. 
From this point of view, the RIXS spectra reflect the charge excitation processes 
in which Cu3$d$ electrons are selectively excited from the lower occupied bands 
to the upper unoccupied bands. 
In Fig.~\ref{Fig:dos}, the total and the Cu3$d$ partial 
density of states (DOS) are shown. 
We find only one DOS weight around $\omega \approx 1$ eV above the chemical potential, 
and four DOS weights around $\omega \approx -2, -4, -6, -9$ eV below the chemical potential. 
We can consider naturally that the three RIXS weights 
around 2, 6 and 10 eV in Fig.~\ref{Fig:1ddpsp} are attributable 
to the excitation processes in which the electrons occupying the bands 
$\omega \approx -2, -6$, and $-9$ eV are excited to the upper unoccupied band. 
Of interest is that the sharp peak in the total DOS 
around $\omega \approx -4$ eV, which corresponds
to the non-bonding oxygen band, is not reflected in the RIXS spectra at all. 
This is because the oxygen non-bonding band does not contain
any components of the Cu3$d$ orbitals and the other orbitals
which are enough localized at the Cu-sites to be coupled strongly to the Cu1$s$ orbitals.
In this sense, the Cu1$s$-4$p$ RIXS measures the Cu3$d$-orbital-selective charge excitations. 
In general, only the bands which are hybridized with the localized orbitals 
strongly coupled to the core hole are observable in the RIXS spectra. 

We should make a distinction between the 2, 6 eV peaks 
and  the 10 eV peak in the RIXS spectra (Fig.~\ref{Fig:1ddpsp}). 
Since the DOS peaks around $\omega \approx 1 $ and $-9$ eV correspond 
respectively to the upper and lower Hubbard bands, 
the RIXS spectral weight around 10 eV in Fig.~\ref{Fig:1ddpsp} 
is related to the charge excitation process 
across the Mott-Hubbard energy gap, which is usually of the order of $U_d$. 
On the other hand, the RIXS spectral weights around 2 and 6 eV 
are related to the charge transfer (CT) process 
in which the Cu3$d$ electrons mixed in the wide O2$p$ bands 
are excited to the upper empty Cu3$d$ Hubbard band. 
This energy scale of the CT excitation is basically connected 
to the CT energy $\varepsilon_{p}-\varepsilon_{d}$. 
We should note that the main spectral weight originates from the CT process, 
and the excitations across the Mott-Hubbard gap provides 
only minor contribution to the RIXS spectra. 
Therefore, the validity of the theoretical analyses 
based on the single-band Hubbard model~\cite{Rf:Tsutsui2000}, 
which consider only Cu sites, is unclear. 

We give some remarks on the present calculations. 
At first, we discuss the validity of the application of the HF
theory to the present system. 
Although the experiments are usually performed above the N\'eel temperature, 
the HF analysis is valid in principle only at the absolute zero temperature. 
In actual low-dimensional systems including the cuprates, 
the fluctuations are very strong to reduce the N\'eel temperature 
and the magnitudes of order parameters. 
However we would like to stress that it is not so important for the RIXS spectra 
whether the AF long range order is attained or only short range AF correlation 
exists. The reasons are as follows. 
(i) The AF transition does not affect the RIXS spectra so drastically, 
since the RIXS spectra basically reflect only the charge sector 
in the scattering process. 
(ii) The RIXS occurs, basically being well localized in the space, 
due to the localized nature of the Cu1$s$ core hole. 
In addition to these points, by noting that 
the AF ground states in Mott insulators 
may be continued adiabatically to the HF AF ground state 
at the absolute zero temperature in the limit of strong Coulomb repulsion, 
we consider that the AF HF theory provides a good starting point 
to the analysis~\cite{Rf:Bulut1994}. 
But we should note here the following point. 
As we have mentioned, the RIXS spectral weight is roughly determined 
by the product of the Cu3$d$ partial occupation numbers 
in the occupied and the unoccupied bands. 
Since the HF theory usually underestimates the magnitude of the localized moment, 
the product of Cu3$d$ electron and hole occupation numbers $n_{i\sigma}(1-n_{i\sigma})$ 
in the local picture is overestimated. 
Thus the intensity around 10 eV in the RIXS spectra is overestimated within the HF theory. 
It is natural to consider that the CT excitations, 
rather than the charge excitation process across the Mott-Hubbard gap, 
provide the main contribution to the spectral weight, 
and the 10 eV peak might not be observable in reality. 

We have considered only the first order terms in the core hole potential $V$. 
Actually we investigated the effects of higher orders in $V$ within 
the $T$-matrix approximation as in usual single impurity problems (Results are not shown). 
However we found that the spectral shape is not so drastically changed 
by the higher orders of $V$, 
although the absolute magnitude of the spectra is changed. 

In the present work, we have analyzed theoretically 
the RIXS spectra in the Q1D cuprates. 
The peak positions and the momentum-transfer dependence of the RIXS spectra 
are explained in a semiquantitative agreement with the experimental results. 
We have shown that the RIXS measurement in cuprates is a unique technique 
to clarify the Cu3$d$-orbital-selective charge excitations. 

The authors are much grateful to Prof. M. Hasan for the communications. 
The numerical computation was partly performed 
at the Yukawa Institute Computer Facility.


\begin{thebibliography}{99}
\bibitem{Rf:Kotani2001} A. Kotani and S. Shin: 
Rev. Mod. Phys. \textbf{73} (2001) 203.

\bibitem{Rf:Kao1996} C. -C. Kao
{\it et al.}: 
Phys. Rev. B \textbf{54} (1996) 16361.

\bibitem{Rf:Hill1998} J. P. Hill
{\it et al.}:
Phys. Rev. Lett. \textbf{80} (1998) 4967.

\bibitem{Rf:Abbamonte1999} P. Abbamonte
{\it et al.}: 
Phys. Rev. Lett. \textbf{83} (1999) 860.

\bibitem{Rf:Hasan2000} M.Z. Hasan
{\it et al.}:
Science \textbf{288} (2000) 1811.

\bibitem{Rf:Hasan2002} M.Z. Hasan
{\it et al.}:
Phys. Rev. Lett. \textbf{88} (2002) 177403.
\label{Rf:Hasan2002}

\bibitem{Rf:Kim2002} Y. J. Kim
{\it et al.}: 
Phys. Rev. Lett. \textbf{89} (2002) 177003.

\bibitem{Rf:Inami2003} T. Inami
{\it et al.}: 
Phys. Rev. B \textbf{67} (2003) 045108.

\bibitem{Rf:Kim2003c} Y. J. Kim
{\it et al.}:
cond-mat/0307497.

\bibitem{Rf:Motoyama1996} N. Motoyama
{\it et al.}:
Phys. Rev. Lett. \textbf{76} (1996) 3212.

\bibitem{Rf:Kojima1997} K. M. Kojima
{\it et al.}:
Phys. Rev. Lett. \textbf{78} (1997) 1787.

\bibitem{Rf:Tsutsui2000} K. Tsutsui
{\it et al.}:
Phys. Rev. Lett. \textbf{61} (2000) 7180.

\bibitem{Rf:Neudert2000} R. Neudert
{\it et al.}:
Phys. Rev. B \textbf{62} (2000) 10752.

\bibitem{Rf:Platzman1998} P. M. Platzman and E. D. Isaacs: 
Phys. Rev. B \textbf{57} (1998) 11107.

\bibitem{Rf:Nozieres1974} P. Nozi\`eres and E. Abrahams: 
Phys. Rev. B \textbf{10} (1974) 3099.

\bibitem{Rf:Bulut1994} N. Bulut
{\it et al.}:
Phys. Rev. Lett. \textbf{73} (1994) 748.

\end{thebibliography}
\end{document}